\documentclass{article}
\usepackage{arxiv}
\usepackage[utf8]{inputenc} 
\usepackage[T1]{fontenc}    
\usepackage[colorlinks=true, allcolors=blue]{hyperref}       
\usepackage{url}            
\usepackage{booktabs}       
\usepackage{amsfonts}       
\usepackage{nicefrac}       
\usepackage{microtype}      
\usepackage{lipsum}		
\usepackage{graphicx}
\usepackage{natbib}
\usepackage{doi}
\usepackage{adjustbox}
\usepackage[version=4]{mhchem}

\title{ALMA detection of acetone, disulfur monoxide, and carbon monoxide in the Jupiter volcanic moon Io}

\author{ {\color{red}Arijit  Manna}\\
	Midnapore City College\\
	Kuturia, Bhadutala, Paschim Medinipur, \\West Bengal, India 721129 \\
	\texttt{Mannaarijit@hotmail.com} \\
	\And
	{\color{red}Sabyasachi Pal} \\
	Indian Centre for Space Physics\\43 Chalantika, Garia Station Road, \\Kolkata, India 700084\\
	Midnapore City College\\
	Kuturia, Bhadutala, Paschim Medinipur, \\West Bengal, India 721129 \\
	\texttt{sabya.pal@gmail.com} \\
}

\renewcommand{\shorttitle}{\ce{CH3COCH3}, \ce{S2O}, and CO in the atmosphere of Io}

\begin{document}
\maketitle

\begin{abstract}
	The extremely thin atmosphere of Jupiter's volcanic moon Io primarily consists of sulfur (S), sodium (Na), and oxygen (O) molecules that are controlled by the combination of the sublimation and volcanic outgasses. We present the first spectroscopic detection of the two rotational emission lines of acetone (\ce{CH3COCH3}) and a single emission line of disulfur monoxide (\ce{S2O}), and carbon monoxide (CO) at frequency $\nu$ = 346.539, 346.667, 346.543, and 345.795 GHz respectively using the archival data of high-resolution Atacama Large Millimeter/Submillimeter Array (ALMA) interferometer with band 7 observation. All molecular species are detected with $\ge$5$\sigma$ statistical significance. The Jupiter's moon Io is the most volcanically active body in the solar system with a very thin and spatially variable atmosphere. The volcanic gas \ce{CH3COCH3}, \ce{S2O}, and CO are mainly coming from volcanic plumes. The statistical column density of \ce{CH3COCH3} line is N(\ce{CH3COCH3}) = 3.18$\times$10$^{15}$ cm$^{-2}$ but for the cases of \ce{S2O} and CO, the column densities are N(\ce{S2O}) = 2.63$\times$10$^{16}$ cm$^{-2}$ and N(CO) = 5.27$\times$10$^{15}$ cm$^{-2}$ respectively. The carbon monoxide gas is mainly formed by the photolysis of the volcanic gas acetone.
\end{abstract}


	\section{Introduction}
In our solar system, the thin atmosphere of Jupiter's volcanic moon Io primarily consists of various volatile substances like \ce{SO2}, SO, NaCl, and the low amount of \ce{H2O} \citep{lel07}. The volcanic species in moon Io are mainly formed by the combination of specific sublimation of volcanic frost layer and outgassing \citep{lel07}. 

The absorption band of \ce{SO2} in the thin atmosphere of Io was first detected in Infrared wavelength \citep{ts11}. The first rotational emission lines of NaCl were detected in the atmosphere of Io using submillimeter observations \citep{lel03}. \citet{lel03} proposed that NaCl is formed with the mixing of sulfur and oxygen compounds in the volcanic atmosphere. The volcanic atmosphere of Io also consists of S, \ce{S2}, SO, and a small number of potassium compounds \citep{lel03}. The kinetic and thermodynamic calculation gives evidence that the volcanic gases are in chemical equilibrium on Io, as on Earth, when they erupt \citep{Zol98}. The volcanic gas \ce{SO2} and SO are extremely unstable and reactive. In the volcanic atmosphere of Io, SO gas is decomposed by the disproportionation method and should form \ce{SO2} and \ce{S2O} gases \citep{ste03}. \\
\begin{center}
	3\ce{SO}$\rightarrow$\ce{SO2}+\ce{S2O} ------ (1)\\
\end{center}
The disulfur monoxide (\ce{S2O}) is primarily known to form as red deposits when condensed at low temperatures. The \ce{S2O} molecular gas is responsible for the red feature which is observed in some volcanoes on Io. The spectrum of \ce{S2O} between 200 and 550 nm shows the characterized absorption feature in the thin atmosphere of Io \citep{hap89}.
\begin{figure*}
	\centering
	\scriptsize
	\includegraphics[width=0.5\textwidth]{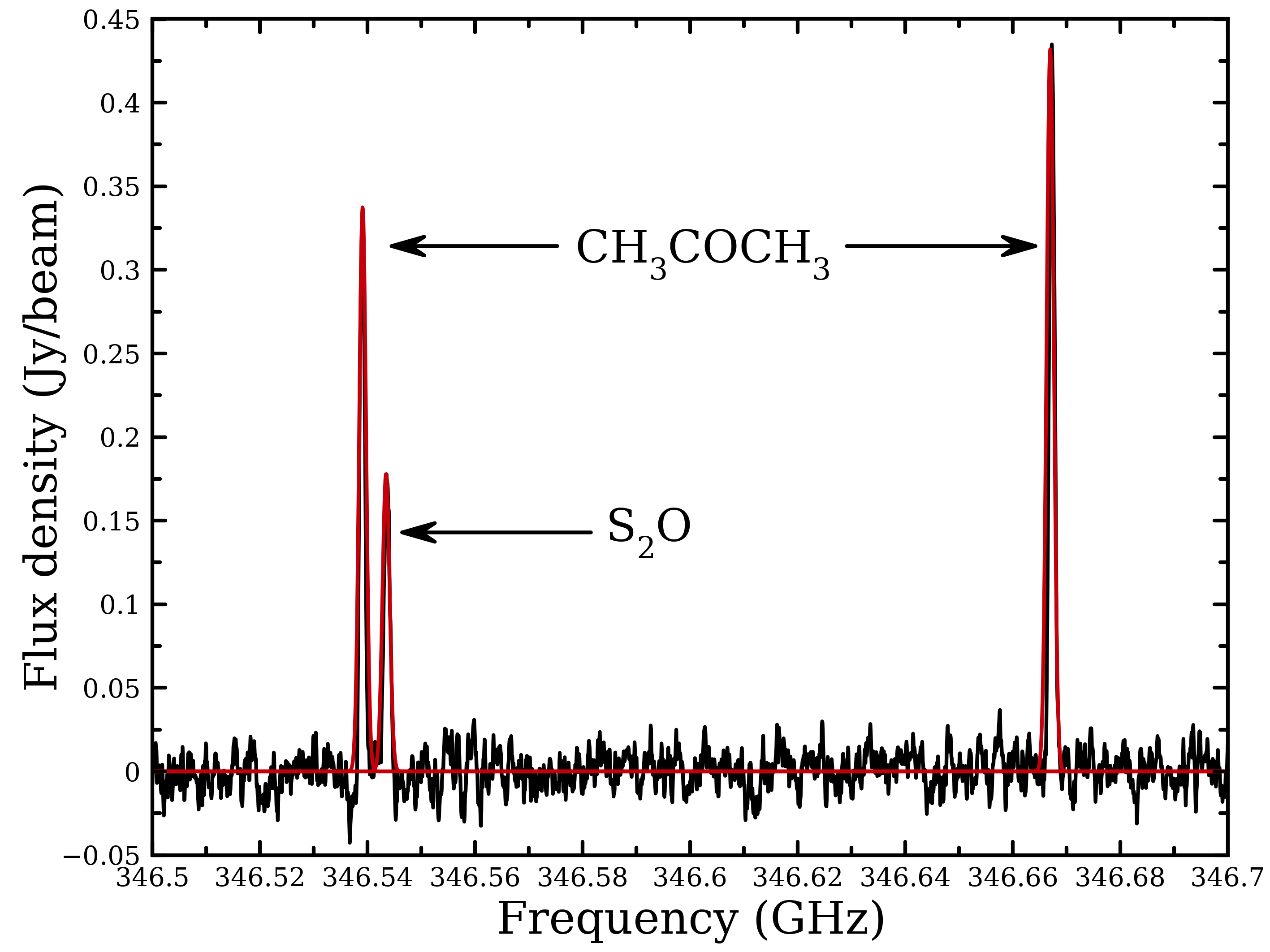}\includegraphics[width=0.5\textwidth]{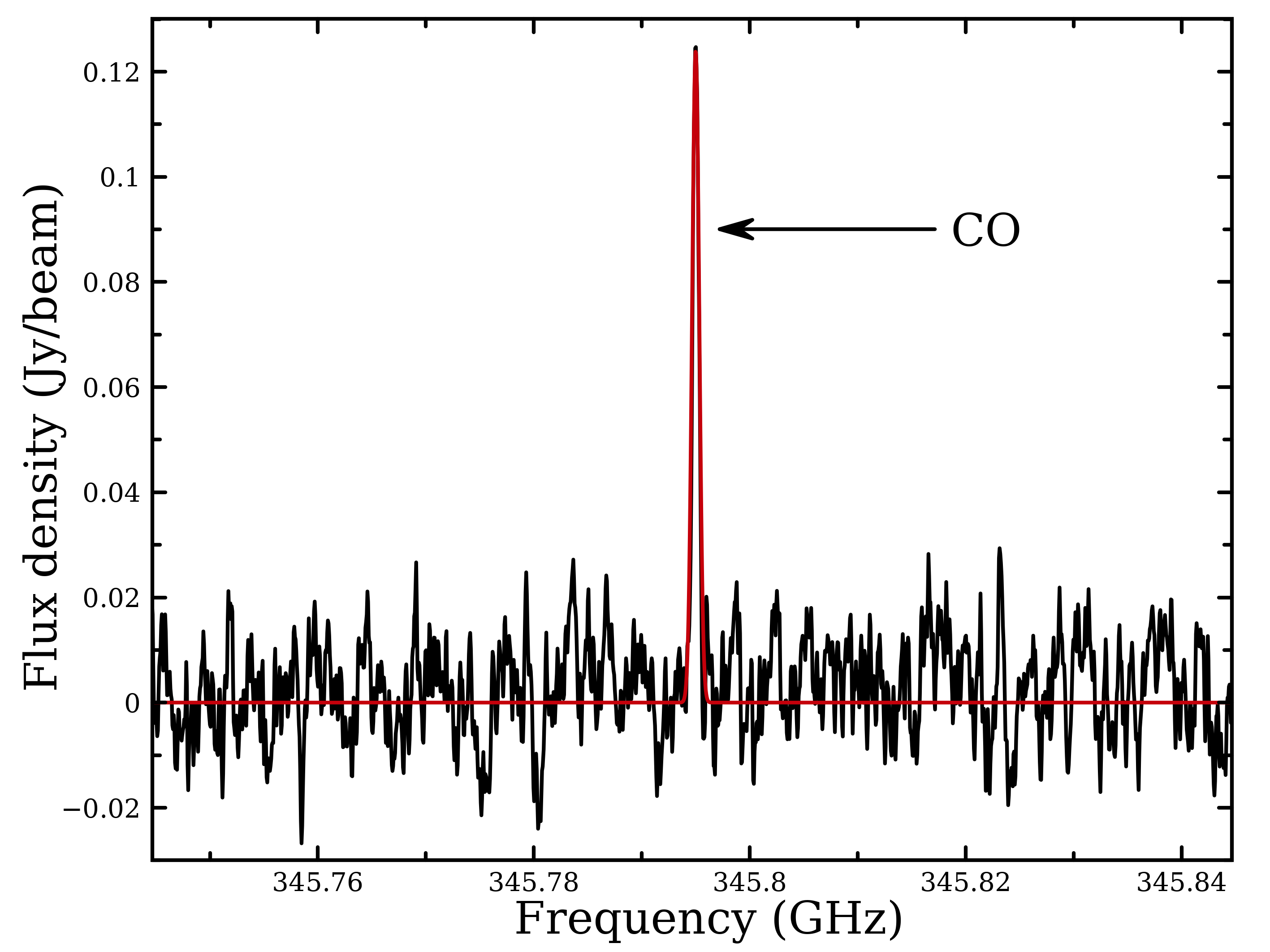}
	\caption{The rotational molecular emission spectrums of \ce{CH3COCH3}, \ce{S2O}, and CO at frequency $\nu$ = 346.539, 346.667, 346.543, and 345.795 GHz respectively in the volcanic moon Io using ALMA band 7 observation. The rotational emission spectrum of volatile species in the atmosphere of Io was made by reducing ALMA data cubes from the center of the Io spectral map within the boundary box region of 1.3$^{\prime\prime}\times$1.6$^{\prime\prime}$. The solid red gaussian feature corresponds to the best fit model of column density.}
	\label{fig:emission} 
\end{figure*}

The volcanic gas acetone is known as the simplest aliphatic ketone which is mostly found in the global atmosphere via the emission from biomass burning and oxidation of non-methane hydrocarbon (NMHCs) \citep{Sin94}. \citet{Sin95} found a sufficient amount of acetone in the global upper troposphere. The understanding of the formation mechanism of acetone in the global atmosphere is very critical because the photochemistry of acetone plays a big role in the production of odd-hydrogen radical (HO$_{x}$) \citep{mck97}. In the volcanic atmosphere, the ketone species acetone can be removed by the reaction with either photolysis or hydroxyl radical (OH). In the upper troposphere, the reaction of acetone with OH radical is very important and the removal process of acetone via reaction with OH radical becomes very faster below 5 km altitude \citep{gie98}. The atmospheric photodissociation of acetone can proceed by the following two reactions \\
\begin{center}
	\ce{CH3COCH3}+h$\nu\rightarrow$2\ce{CH3}+CO ------ (2)\\
\end{center}
\begin{center}
	\ce{CH3COCH3}+h$\nu\rightarrow$\ce{CH3}+\ce{CH3CO} --- (3)\\
\end{center}
In the solar system, acetone is the second detected molecule with ten atoms after glycine. Glycine is the first ten atoms molecule detected in the solar system which is observed in the atmosphere of Venus between the frequency range of $\nu$ = 245--262 GHz with $\geq3\sigma$ statistical significance using the ALMA \citep{Man20}. Earlier, acetone was first detected in the hot molecular core of Sgr B2 \citep{Com87, Sny02}. 

\begin{table*}
	\caption{Molecular and spectral properties of \ce{CH3COCH3}, \ce{S2O}, and CO in the atmosphere of Io .}
	\centering
	\scriptsize
	\begin{tabular}{|c|c|c|c|c|c|c|c|c|c|}
		\hline \hline
		Species&Frequency&Transition&Lower&$\log$(A$_{ij}$)&Flux density& Line area&FWHM\\
		& (GHz)    & (J)     &energy (K)&&(Jy beam$^{-1}$)&(Jy beam$^{-1}$$\times$km s$^{-1}$)&(km s$^{-1}$)\\	
		\hline		
		\ce{CH3COCH3}&346.539&$26_{(19,6)}-25_{(14,11)}AE$&~261.136&--5.950&~0.34&0.365$\pm$0.005&0.973$\pm$0.005\\
		&346.667&$55_{(15,40)}-55_{(14,41)}EA$&1015.317&--3.214&~0.46&0.479$\pm$0.004&1.014$\pm$0.009\\
		\ce{S2O}&346.543&$38_{(0,38)}-37_{(1,37)}$&~216.650&--3.513&~0.17&0.212$\pm$0.003&1.102$\pm$0.018\\
		CO&345.795&3--2&~~11.5350&--5.602&~0.13&0.048$\pm$0.001&0.414$\pm$0.001\\
		\hline
	\end{tabular}
	\label{tab:MOLECULAR DATA}
\end{table*}

The hydrogen gas is depleted in the volcanic atmosphere of Io from \ce{H2O}, \ce{H2S} and \ce{H2} which is not expected in the thin atmosphere of Io \citep{Feg20}. \citet{Sch04} first presented the theoretical simulation of the existence of carbon compounds like \ce{CO2}, CO, and OCS in the volcanic atmosphere of Io. The Voyager spacecraft find the first upper limit of carbon-based gases in the Loki plume \citep{Pea79}. The carbon and sulfur mixing ratio in the Loki plume is very small than the volcanic gases in the Pele plume and other hot spots. \citet{Sch04} calculated the kinetic and thermal equilibrium equation to find the carbon chemistry in the atmosphere of Io. 

In this article, we present the first spectroscopic detections of \ce{CH3COCH3}, \ce{S2O}, and CO at $\nu$ = 346.539, 346.667, 346.543, and 345.795 GHz in the atmosphere of Jupiter's volcanic moon Io using ALMA band 7 data with a 12m array. In Sect.~\ref{sec:obs}, we discuss the observations and data reductions. The result and discussion of the detection of volcanic species are shown in Sect.~\ref{sec:result}. The summary is presented in Sect.~\ref{sec:discussion}.

\section{Observations and data reduction}
\label{sec:obs} 
We used the high-resolution interferometric data of Io from the ALMA\footnote{\href{https://almascience.nao.ac.jp/asax/}{https://almascience.nao.ac.jp/asax/}} archive with band 7 observation. The observation of Io in band 7 begins at 9h8m40.1s UTC on October 17, 2012, and ends at 10h36m02.7s UTC on October 17, 2012. The device was set up in four spectral windows with a frequency range of 344.99--347.34 GHz. The XX, YY, and XY-type signal correlators were used with an integration time of 1572.48 seconds. A total of 21 antennas was operational during the observation. J0538--440 was used as a bandpass, and J044907+112125 was used as a phase calibrator. During observation, the distance between Earth and Io was 4.3627 AU and the angular size of Io was 1.15$\prime\prime$.

For initial data calibration and imaging, we used the Common Astronomy Software Application ({\tt CASA 5.4.1})\footnote{\href{https://casaguides.nrao.edu/}{https://casaguides.nrao.edu/}}. The continuum flux density for each baseline was scaled and matched with Butler-JPL-Horizons 2012 \citep{But12} for the Io flux model with 5\% accuracy. After the flux and bandpass calibration with flagging bad data, we split out the target data for continuum and spectral imaging using task {\tt mstransform} with rest frequencies in each spectral window. The continuum subtraction was done by task {\tt uvcontsub}. After spectral imaging, we found the disk average spectrum of high abundant two rotational emission lines of volcanic gas \ce{CH3COCH3} and single rotational emission lines of \ce{S2O} and CO. The final rotational molecular spectrum was corrected for doppler-shift to Io topocentric spectral frame using NASA JPL Horizons Topocentric radial velocity.

\section{Result and discussion}
\label{sec:result} 
\subsection{Rotational molecular lines of \ce{CH3COCH3}, \ce{S2O}, and CO in the atmosphere of Io}
We detected the high abundant two rotational emission lines of volcanic gas \ce{CH3COCH3} at $\nu$ = 346.539, 346.667 and single rotational emission lines of \ce{S2O} and CO at $\nu$ = 346.543 GHz and 345.795 GHz respectively in the atmosphere of the volcanic moon Io. The detected molecular spectrum in the atmosphere of Io with gaussian fitting is shown in Fig.~\ref{fig:emission}. The molecular properties of detected species and spectral fitting parameters are shown in Tab.~\ref{tab:MOLECULAR DATA}.  The spectral peaks were identified using the online Splatalogue\footnote{\href{https://splatalogue.online//}{https://splatalogue.online//}} database for astronomical molecular spectroscopy and also using Cologne Database for Molecular Spectroscopy (CDMS)\footnote{\href{https://cdms.astro.uni-koeln.de/cgi-bin/cdmssearch}{https://cdms.astro.uni-koeln.de/cgi-bin/cdmssearch}} \citep{Mu01}. The acetone (\ce{CH3COCH3}) is another volcanic gas that was found first time in the atmosphere of Io. The volcanic gas acetone is also found on Earth-based volcanos \citep{pie18}. In the volcanic atmosphere of Io, another volcanic species \ce{S2O} is detected for the first time using ALMA band 7 observation with a single transition line at 346.543 GHz. We also detected the rotational emission line of CO (J = 3--2) at frequency $\nu$ = 345.795 GHz from the completely unpredictable atmosphere of Io. Carbon monoxide gas is rare species in the volcanic moon. We propose that the CO (J=3--2) gas were formed in the volcanic atmosphere due to photolysis of acetone (\ce{CH3COCH3} + $h \nu \rightarrow$CO+\ce{CH4}).

\subsection{Radiative transfer modeling of detected volcanic gas}
We used the {\tt XCLASS}\footnote{\href{https://xclass.astro.uni-koeln.de/}{https://xclass.astro.uni-koeln.de/}} software package in {\tt CASA} for radiative transfer simulation of the emission line of \ce{CH3COCH3}, \ce{S2O}, and CO. The spectroscopic parameters were taken from CDMS \citep{Mu01} and molecular spectroscopic database of the Jet Propulsion Laboratory \citep{pic98}. In the version of {\tt XCLASS} used in this study, a global linear fit in log-log space is performed for each species on the tabulated values of the partition function between 9.375 and 300 K. The modeling was carried out on a species-by-species basis, with a species being one molecule or one vibrationally excited state of a molecule. During the radiative transfer simulation using the {\tt XCLASS} package, we initially assumed that the level populations of each species were described by a single excitation temperature, $T_{rot}$. This statement holds true in two situations: at high density, where collisions are regular enough to support the local thermodynamic equilibrium (LTE) approximation, and at very low density, where $T_{rot}$ equals the temperature of the cosmic microwave background, $T_{CMB}$ = 2.73 K.

To extract the abundance of detected species in the atmosphere of Io, we calculated the statistical column density of \ce{CH3COCH3}, \ce{S2O} and CO and compared them with the most abundant volcanic gas \ce{SO2}. The beam averaged column density of \ce{CH3COCH3} is N(\ce{CH3COCH3}) = 3.18$\times$10$^{15}$ cm$^{-2}$ and for the case of \ce{S2O} and CO, the column densities were N(\ce{S2O}) = 2.63$\times$10$^{16}$ cm$^{-2}$ and N(CO) = 5.27$\times$10$^{15}$ cm$^{-2}$ respectively. We assumed the volcanic species \ce{CH3COCH3} and \ce{S2O} were colocated and share the same kinetic temperature. In volcanic atmosphere, \ce{S2O} is primarily derived from the sublimation of volcanic out-gases. We take the column density of \ce{SO2} as 1.5$\times$10$^{16}$ cm$^{-2}$ \citep{pat20} to calculate the abundance of detected species in the moon Io, which is enlisted in Tab.~\ref{tab:abun}. 

\begin{table}
	\centering
	\caption{Abundance of detected species in volcanic moon Io.}
	\begin{tabular}{|c|c|c|c|c|c|c|c|c|c|c|}
		\hline \hline
		Species&T$_{rot}$&Column density&Abundance with \\
		& (K)        &(cm$^{-2}$)&respect to \ce{SO2}\\
		\hline
		\ce{CH3COCH3}&30&3.18$\times$10$^{15}$&0.21\\
		\ce{S2O}&20&2.63$\times$10$^{16}$&1.75\\
		CO&15&5.27$\times$10$^{15}$&0.35\\
		\hline
	\end{tabular}
	\label{tab:abun}
\end{table}

\section{Summery}
\label{sec:discussion} 
In this article, we present the first spectroscopic detection of the rotational molecular lines of \ce{CH3COCH3}, \ce{S2O}, and CO at frequency $\nu$ = 346.539, 346.667, 346.543, and 345.795 GHz respectively in the atmosphere of Jupiter's volcanic moon Io with ALMA band 7 observation. The formation mechanism of \ce{CH3COCH3} in the volcanic atmosphere of Io is completely unknown but the spectroscopic detection implies that a big amount of methane compounds may exist. The CO gas at 345.795 GHz was detected first time on Io. We proposed that the CO gas were formed with the photolysis of acetone (\ce{CH3COCH3} + $h \nu \rightarrow$CO+\ce{CH4}). The volatile gas acetone and carbon monoxide in the atmosphere of Io were probably rising in the brightness of the Loki hot spot. The detected species in the volcanic atmosphere using ALMA calls for more comprehensive studies of their rotational lines and requires a re-examination of whether they co-exist in the same atmospheric position to get more specific ways to predict the chemical formation process.


\section*{Acknowledgement}
This paper makes use of the following ALMA data: ADS /JAO.ALMA\#2011.0.00779.S. ALMA is a partnership of ESO (representing its member states), NSF (USA), and NINS (Japan), together with NRC (Canada), MOST and ASIAA (Taiwan), and KASI (Republic of Korea), in cooperation with the Republic of Chile. The Joint ALMA Observatory is operated by ESO, AUI/NRAO, and NAOJ. The data that support the plots within this paper and other findings of this study are available from the corresponding author upon reasonable request. The raw ALMA data are publicly
available at \href{https://almascience.nao.ac.jp/asax/}{https://almascience.nao.ac.jp/asax/}.
\section*{Data Availability Statement}
The data that support the plots within this paper and other findings of this study are available from the corresponding author upon reasonable request. The raw ALMA data are publicly
available at \href{https://almascience.nao.ac.jp/asax/}{https://almascience.nao.ac.jp/asax/} (project id : 2011.0.00779.S.).

\end{document}